\documentclass[a4paper,UKenglish]{lipics-v2018}

\usepackage{microtype}

\title{Padovan Heaps}

\author{Vladan Majerech}{Department of Theoretical Computer Science and Mathematical Logic (KTIML), Charles University, Malostransk\'e n\'am\v est\'\i\ 25, Prague 118 00, Czech Republic}{maj@ktiml.mff.cuni.cz}{ https://orcid.org/0000-0003-3006-2002}{}

\authorrunning{V. Majerech}

\Copyright{Vladan Majerech}

\subjclass{Information systems$\rightarrow$Information storage systems$\rightarrow$Record storage systems$\rightarrow$Record storage alternatives$\rightarrow$Heap (data structure)}


\category{}

\relatedversion{}

\supplement{}

\funding{}

\acknowledgements{}

\EventEditors{}
\EventNoEds{1}
\EventLongTitle{}
\EventShortTitle{}
\EventAcronym{}
\EventYear{2019}
\EventDate{}
\EventLocation{}
\EventLogo{}
\SeriesVolume{}
\ArticleNo{1}
\nolinenumbers 
\hideLIPIcs  

\begin{document}
\newcommand{\FindMin}{{\sl FindMin\/}}
\newcommand{\DeleteMin}{{\sl DeleteMin\/}}
\newcommand{\Delete}{{\sl Delete\/}}
\newcommand{\Meld}{{\sl Meld\/}}
\newcommand{\Cut}{{\sl Cut\/}}
\newcommand{\Insert}{{\sl Insert\/}}
\newcommand{\MakeSafe}{{\sl MakeSafe\/}}
\newcommand{\Decrement}{{\sl DecreaseKey\/}}
\newcommand{\Right}{\hbox{\sl right\/}}
\newcommand{\Left}{\hbox{\sl left\/}}
\newcommand{\nill}{\hbox{\sl null\/}}
\newcommand{\Parent}{{\sl parent\/}}
\newcommand{\numfootnote}{\footnote}

\maketitle

\begin{abstract}
We analyze priority queues of Fibonacci family. The paper is inspired by Violation heap \cite{ViolationHeaps}, where A. Elmasry saves one pointer in representation of Fibonacci heap nodes while achieving the same amortized bounds as Fibonacci heaps \cite{FibonacciHeaps} of M. L. Fredman and R. E. Tarjan. Unfortunately author forces the heaps to be wide, what goes against optimal heap principles. Our goal is to achieve the same result, but with much narrower heaps. We follow the principle of superexpensive comparison so we try to remember results of all comparisons and never compare elements that cannot be minimal. We delay comparisons as long as possible. Actually I have always want to share superexpensive comparison principle ideas, discovery of Padovan heaps allowed me to do so. Of course saving one pointer is not that big goal, but I hope the presented reasoning and amortized analysis of the resulting heaps is worth a publication.
\end{abstract}

\section{Analysis of heaps based on superexpensive comparison principle}
Heaps are data structures supporting \Insert, \FindMin\ and \DeleteMin\ operations\numfootnote{All heaps mentioned in this article support naturally \Meld\ operation.}. Heaps of Fibonacci family \cite{FibonacciHeaps} support \Decrement\ operation as well\numfootnote{and general \Delete\ with $O(\log n)$ complexity}.
Let $n$ denote number of elements represented by the heap. Our goal is to achieve amortized analysis with complexity of \DeleteMin\ logarithmic ($O(\log n)$) and complexities of remaining operations constant ($O(1)$). Both Binomial and Fibonacci families of heaps have these properties, but Binomial heaps don't support \Decrement\ operation.

Both Binomial and Fibonacci family heaps represent heap elements as vertices in directed forest of heap-ordered trees (key in a child is no less than key in it's parent). Roots of the trees represent candidates for minimum. Internal representation of the forest will be discussed later.

Superexpensive comparison principle leads us to implement \Insert\ in constant time by just adding an isolated vertex to the forest (standard implementation compares the element key with current minimum and updates minimum, what increases required time by a constant; the comparison result is not remembered as a graph edge).

The basis of both these families of heaps is the \FindMin\ operation\numfootnote{Thanks to caching minimum, this is called just after \DeleteMin\ in standard implementation.}. Its worst case time is $O(n)$, but this time could be prepaid by preceeding operations. Each comparison would result in creating an edge between current tree roots, therefore reducing number of trees by $1$. \FindMin\ finishes with just one tree (unless we violate superexpensive comparison principle). Potential $\Phi_0$ equal to number of trees $\tau$ ($\Phi_0=\tau$) could pay for the comparisons and maintaining $\Phi_0$ would raise cost of inserts just by a constant.

With appropriate internal representation, \DeleteMin\ operation (preceeded immediately by \FindMin) could be implemented in $O(1)$ worst case time removing current minimum and all incident edges. 
The number of trees in resulting graph would increase by the degree of the minimum minus $1$ so the amortized cost of \DeleteMin\ with respect to $\Phi_0$ is degree of the minimum. This is incentive for maintaining the trees as narrow as possible. 

To achieve small degrees of vertices, ranks and rank invariant were introduced. 
Each vertex $v$ gets nonnegative rank $r_v$. 
Rank invariant stays that the size of a vertex $v$ subtree must be at least exponential in the vertex rank ($\ge c\beta^{r_v}$ for a fixed $c>0$, and $2\ge \beta>1$). 
Rank invariant guarantees the maximal rank is logarithmic ($\lfloor\log_\beta (n/c)\rfloor$).
Standard variants of Binomial and Fibonacci heaps maintain number of vertex children equal to its rank, but in that case they must forget results of some comparisons, what is against superexpensive comparison principle. 
The other option is to allow a number of children to differ from ranks. 
Number of $v$'s children $\iota_v$ smaller than rank $r_v$ would not be a problem, but for $\iota_v$ higher than rank we would need the differences to be prepaid in potential. 
We introduce $\Phi_1$ equal to sum of positive differences between a number of children and rank of a vertex ($\Phi_1=\sum_{v\mid \iota_v>r_v} \iota_v-r_v$).

Thanks to the rank invariant for a minimum $m$, \DeleteMin\ increases $\Phi_0$ by at most logarithm plus decrease of $\Phi_1$ casued by difference of minimum's number of children and rank ($\Delta \Phi_0 = (\iota_m-1) = -1+r_m+(\iota_m-r_m) \le -1+r_m-\Delta\Phi_1 \le \lfloor\log_\beta (n/c)\rfloor - 1 - \Delta\Phi_1$).

To finish the description of Binomial heap operations, we should show how ranks are defined, and how \Insert\ and \FindMin\ are implemented with respect to ranks.  

Rank of a new isolated vertex created by \Insert\ is initialised to $0$.
Comparison of two roots of the same rank $r$ results in edge joining the trees increasing rank of it's root by $1$.
If the sizes of the original trees were at least $c\beta^r$, size of resulting tree is at least $2c\beta^r\ge c\beta^{r+1}$ so this operation preserves rank invariant. 
\FindMin\ works in two phases.
In the first phase roots of the same rank are detected and pairwise joins of the roots of the same rank are done (details later).
In the second phase standard implemention finds minimal root without creating corresponding edges, while our representation creates edges without changing ranks.
Standard impementation of \FindMin\ ends with at most logarithmic number of tree roots of different ranks, our implementation increases $\Phi_1$ by at most logarithm instead.

Potential $\Phi_0$ is enough to pay for the first phase of \FindMin.
The second phase time is bounded by both logarithm and the number of trees prior to the \FindMin.
Standard implementation maintains pointer to the minimum after each operation.
Therefore it calls \FindMin\ just after a \DeleteMin\ and lets \DeleteMin\ pay for the second phase.
According to the superexpensive comparison principle we cannot maintain minimum between user calls of \FindMin\ in our implementation. Instead we introduce potential $\Phi_2$ equal to minimum of logarithm and number of trees ($\Phi_2=\min\{\tau, 1+\lfloor\log_\beta (n/c)\rfloor\}$).
The potential $\Phi_2$ pays for the second phase of \FindMin.
Maintaining $\Phi_2$ increases cost of \DeleteMin\ by at most logarithm ($1+\lfloor\log_\beta (n/c)\rfloor$), and cost of \Insert\ by at most constant ($O(1)$).

When there is no \Decrement, and we increment rank just when roots of the same size are joined, rank invariant for $c=1$ and $\beta=2$ holds (completing analysis of Binomial heaps).

During \Decrement\ the edge from affected vertex to its parent is removed and its parent rank is decremented.
But this would not suffice to maintain the rank invariant for other vertices on the path to the root. 
Standard implementation of \Decrement\ uses cut delaying strategy. Rank of any vertex could be decremented once making vertex critical. Second decrement results in making vervex noncritical, cut and rank decrement of its parent.
According to the superexpensive comparison principle we cannot remove the edge, so we make vervex noncritical and instead of cut we just decrement the rank of its parent (what increments $\Phi_1$). 
So standard cascading cuts are transformed to cascading rank decrements.

Criticality is important just for nonroot vertices. When \FindMin\ connects two tree roots, it makes the child noncritical.
Let noncritical rank $\rho_v$ corresponds to rank ($r_v$) for a noncritical vertex $v$ and is one higher than rank ($r_v+1$) for a critical vertex $v$.

To finish the analysis we have to show two facts. The first is constant cost of \Decrement, the other is that the rank invariant persists.

Let $\Phi_3$ be the number of critical nonroots ($\Phi_3=\bigl|\{v|v{\rm\ has\ parent} \wedge v{\rm \ is\ critical}\}\bigr|$). 
Vertices become critical only during \Decrement.
At most one vertex becomes critical during it and number of rank updates corresponds to number of vertices whose change from critical to noncritical, so \Decrement\ is paid from $\Phi_3$. Actually $\Phi_3$ must also pay to potentials $\Phi_0$ or $\Phi_1$ (it pays to $\Phi_0$ in standard implementation and to $\Phi_1$ in our instead).

We could number the children by order of joining.
Just guarantors of the rank, therefore children whose joining increased rank and whose were not reverted from critical to noncritical are considered. 
At the time of $i$-th join, the rank of the vertex must be at least $i-1$ so the $i$-th child has the same noncritical rank. 
The rank of the $i$-th child could decrease from noncritical rank $i-1$ to $i-2$ iff the child becomes critical.
Let $M_r$ be minimal possible size of the ancestors tree of vertex with rank $r$. Than $M_r=M_{r-2}+\cdots+M_0+M_0+1$.
For $M_{r+1}$ the sum increases by $M_{r-1}$, therefore we get $M_{r+1}=M_r+M_{r-1}$ what leads to $\beta^2=\beta+1$ giving golden ratio $\beta=q=\frac{1}{2}(1+\sqrt{5})\approx 1.618034$ and close connection to Fibonacci sequence, that named the heaps.

The main goal of \cite{ViolationHeaps} is to reduce the size of a heap representation.
This is achieved by dividing vertices to active and inactive.
Only active vertices have guarenteed constant access time to their parents.
This leads to modification of rank invariant when only connected subtrees of active vertices are considered and the size of vertex $v$ and subtree of its active descendants is bounded by $c\beta^{r_v}$.
In \cite{ViolationHeaps} only last two children of a vertex are active.
We would consider at most last two children of a vertex to be active as well.
The order of children is important.
Removal of active children of a vertex of rank $r$ could result in an isolated vertex, therefore cut of the vertex could result in reduction of the vertex rank by $\Theta(\log n)$.
That would be incompatible with $\Phi_1$ used in our analysis of Fibonacci heaps.
This is why we should change the analysis and implementation slightly.

We introduce the third kind of children: Let us call children who are noncritical or critical inner and introduce outer children\numfootnote{Introducing outer vertices for Fibonacci Heaps would make their description cleaner, otherwise it suffices to prevent decrementing rank under 0.}.
For a vertex $v$ let $o_v$, $c_v$, resp. $n_v$ denote number of outer, critical, resp. noncritical children of $v$.

First phase of \FindMin\ creates noncritical children, second phase of \FindMin\ creates outer children. 
Outer children of a vertex would be maintained on the start of its children list. 
They will be followed by inner children implicitly ordered by increasing noncritical rank.
Potential $\Phi_1=\sum_v o_v$ would be number of all outer children.
First change in decrement is that instead of cascading rank decrement we do cascading rank recomputation and it would be implicitly stopped at outer vertices.
Another change is when critical vertex rank is decremented, we create outer vertex instead of creating noncritical one and we move the outer vertex out from it's original position among children to the first place of children list\numfootnote{At least we mark the vertex and do it at appropriate time.}.
Last change in cascading rank recomputation is that only when rank of one of the last two children changes the rank of vertex should be recomputed. 
If the rank drops by more than one, it's considered as (at least) two decrements so even noncritical vertex becomes outer.
To pay for bigger rank decrease during the recomputation, potential $\Phi_4$ acumulating delayed rank decreases should be introduced 
($\Phi_4=\sum_v r_v-c_v-n_v$). We have to prove that $(*)$ $r_v\ge c_v+n_v$ holds all the time.

We have to redefine ranks in a way rank invariant would hold even on active subtrees. 
If the two highest noncritical ranks of inner children differ by at most 1, we let them both be active and define rank of the parent to be one higher than the maximal one.
Otherwise if there is an inner child, there must be child $w_0$ with highest noncritical rank $\rho_{w_0}$, we let it the only active child and the rank of its parent be the same. 
But in this case we don't accept the child $w_0$ to be critical, we should make such critical $w_0$ outer and recompute the rank if that happened.
There will be exceptions at small ranks, but it would not affect the recurrence for the minimal active tree size $M_r$ for given rank $r$.

\vbox to 100mm{
\kern100mm
\hbox{\hss\kern 10mm\pdfximage width 12cm {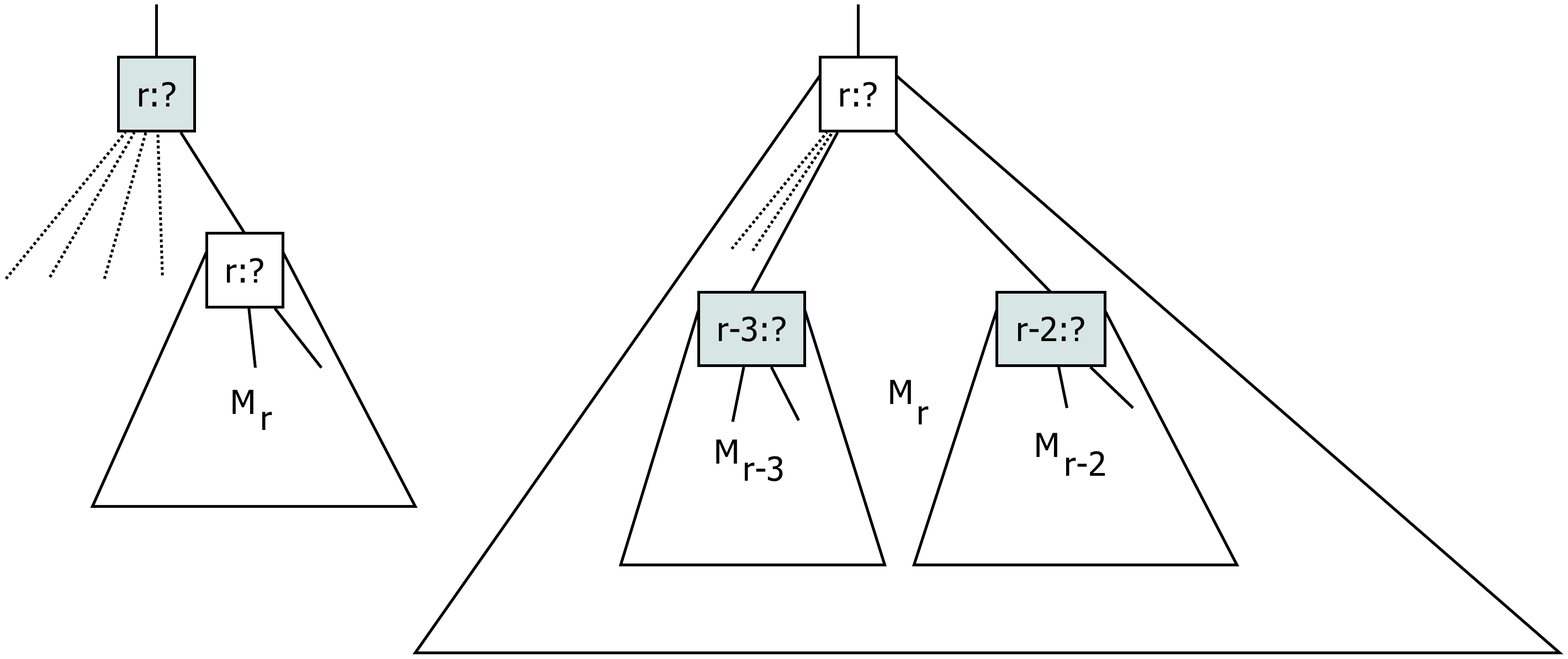}%
\rlap{\smash{\pdfsave
\pdfrefximage\pdflastximage}}\pdfrestore
\kern0cm}
\vss
\hbox{fig 1: Minimal size rekurence; grey color denotes critical child}
\hbox{solid line connects active child to its parent}
\kern5mm
}

If the active child is/becomes critical it's rank is one less than noncritical rank.
So we got $M_r=1+M_{r-2}+M_{r-3}$. It leads to $\beta^3=\beta+1$ giving plastic number $\beta=p={\root3 \of {\frac{1}{2}\bigl(1+\sqrt{23/27}\bigr)}}+{\root3 \of {\frac{1}{2}\bigl(1-\sqrt{23/27}\bigr)}}\approx 1.324718$ and close connection to Padovan sequence, that named the heaps.

We let the rank definition details, proof of invariant $(*)$, and proof the that rank recomputation cost remains constant for Padovan heaps to it's own chapter.
Prior to it, we should mention important implementation details to support the analysis so far.

\section{Implementation details}

\vbox to 65mm{
\hbox{\hss\kern 5mm\pdfximage width 9cm {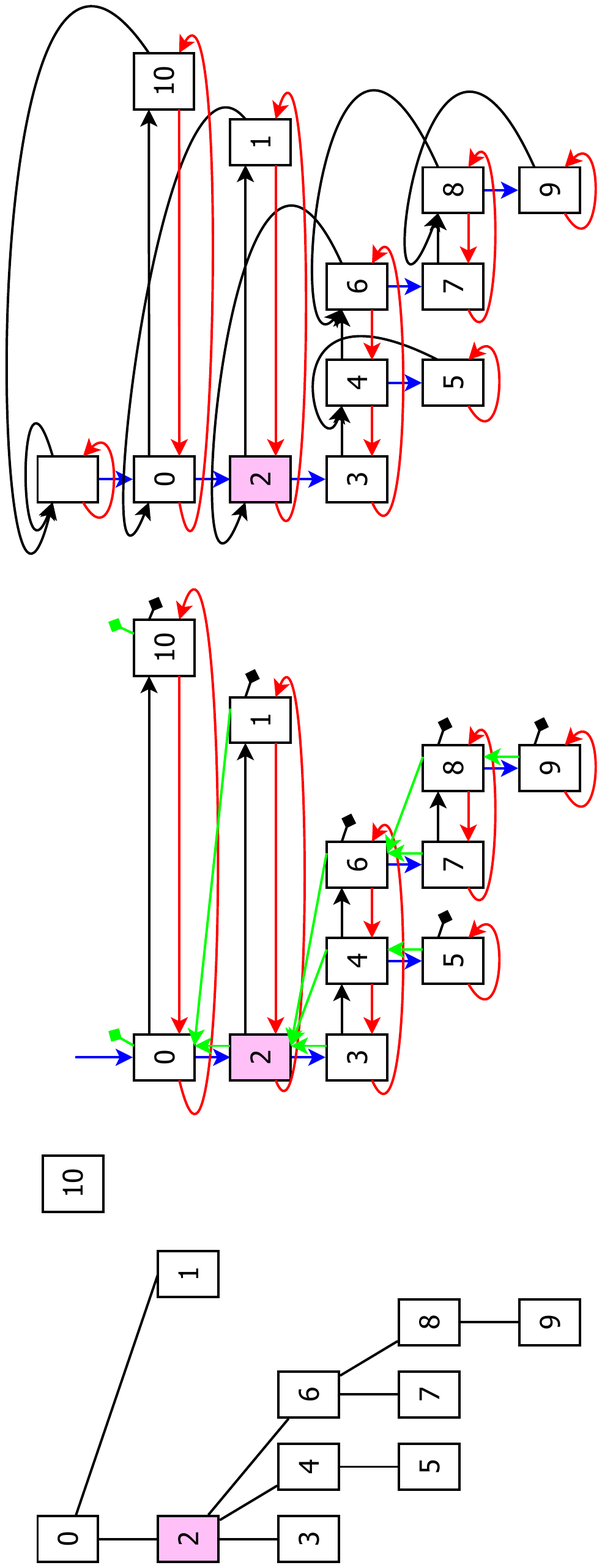}\rlap{\smash{\pdfsave\pdfsetmatrix{0 -1 1 0}\pdfrefximage\pdflastximage}}\pdfrestore
\kern5cm}
\vfil
\hbox{fig 2: forest of heap ordered trees, it's representation in Fibonacci heaps (outer children}
\hbox{variant), and in Padovan heaps; pink color denotes outer child}
\kern5mm
}

Internal representation of the forest is a list of tree roots, each vertex points to the list of its children.
For Decrement interface user should know pointers to vertices.
This is why Insert would return pointer to the vertex for user's future use.
To support Decrement, we require double linked lists. 
I prefere left list to be circular and right ending in \nill\ pointer. 
The circularity of left list gives us access to both ends of the list allowing inserts on either end.
This is not important for Fibonacci heaps, but we will use it in Padovan heaps to insert on diferent ends in first and second phase of \FindMin\ and it will allow us to reinsert children becoming outer to the propper place.

Binomial family heaps neednot represent parent pointers, while Fibonaci family heaps require them for implementation of decrement preventing heap degeneration.
Idea of \cite{ViolationHeaps} is that parent pointers neednot be present at all elements, and in Padovan heaps we save space by saving parent pointer only at right pointer of rightmost lists element.
We can check $v$ is at the end of the list by fact the $v\to\Right\to\Left$ does not point to $v$.  
Accessing parent from elements in the middle of the list would be expensive, but we will not access them in Padovan heap implementation.

We neednot maintain the vertex type information for roots, their corresponding field would be filled arbitrary. \FindMin\ sets the information appropriately when creating the edge.

\section{Details of \FindMin\ implementation}
First phase of \FindMin\ is usually implemented on RAM\numfootnote{The same could be implemented on pointer machine model (when we cannot allocate nonconstant array).
Ranks could be implemented by pointers to global (expandnig when needed) list of integers. 
The list could store placeholders for pointers back to the roots of current \FindMin\ procedure.}
using long enough array adressed by tree root ranks.
Roots are put to empty places in array and if the place is already occupied, the root with same rank is identified and the pair could be compared and joined together according to the result.
This leaves the place empty and increases rank of the resulting root.
After all roots are put to empty places roots from nonempty places in the array are taken to form new list of roots.

The final step is inefective especially when the array is almost empty.
When we have roots in double linked list\numfootnote{Alternatively we could maintain double linked list of occupied places in the array, what would work even when children are not double linked (Binomial heaps).
A stack of at least once used places would work as well.
Its use would be charged to the first phase.}, we could change the procedure slightly.
We don't remove the roots from the list when putting pointer to them to the array.
We remove a root from the list only during join when it becomes a child\numfootnote{We must be careful in list of roots traversal to save pointer to the current root$\to\Right$ before the root could be removed from the list.}.
This implementation detail guarantees we have list of roots of different ranks after the list is traversed.
We could clean the array by traversing the resulting list.
Therefore we visit only nonempty places of the array and the time is bounded by the number of roots at the begining of the second phase of \FindMin, what is bounded both by the number of roots at the begining of the first phase of \FindMin\ and by $1+\lfloor\log_\beta(n/c)\rfloor$.

The second phase of \FindMin\ links the last root with the second last according to the comparisson.
Then it links third last with fourth last\numfootnote{They become second last and third last after the first link.}, and so on, during the cyclic traversal. The second phase ends when there is only one root on the list.

\section{Analysis details for Padovan heaps}
\vbox to 90mm{
\kern 225mm
\hbox{\hss\kern -10mm\pdfximage width 16cm {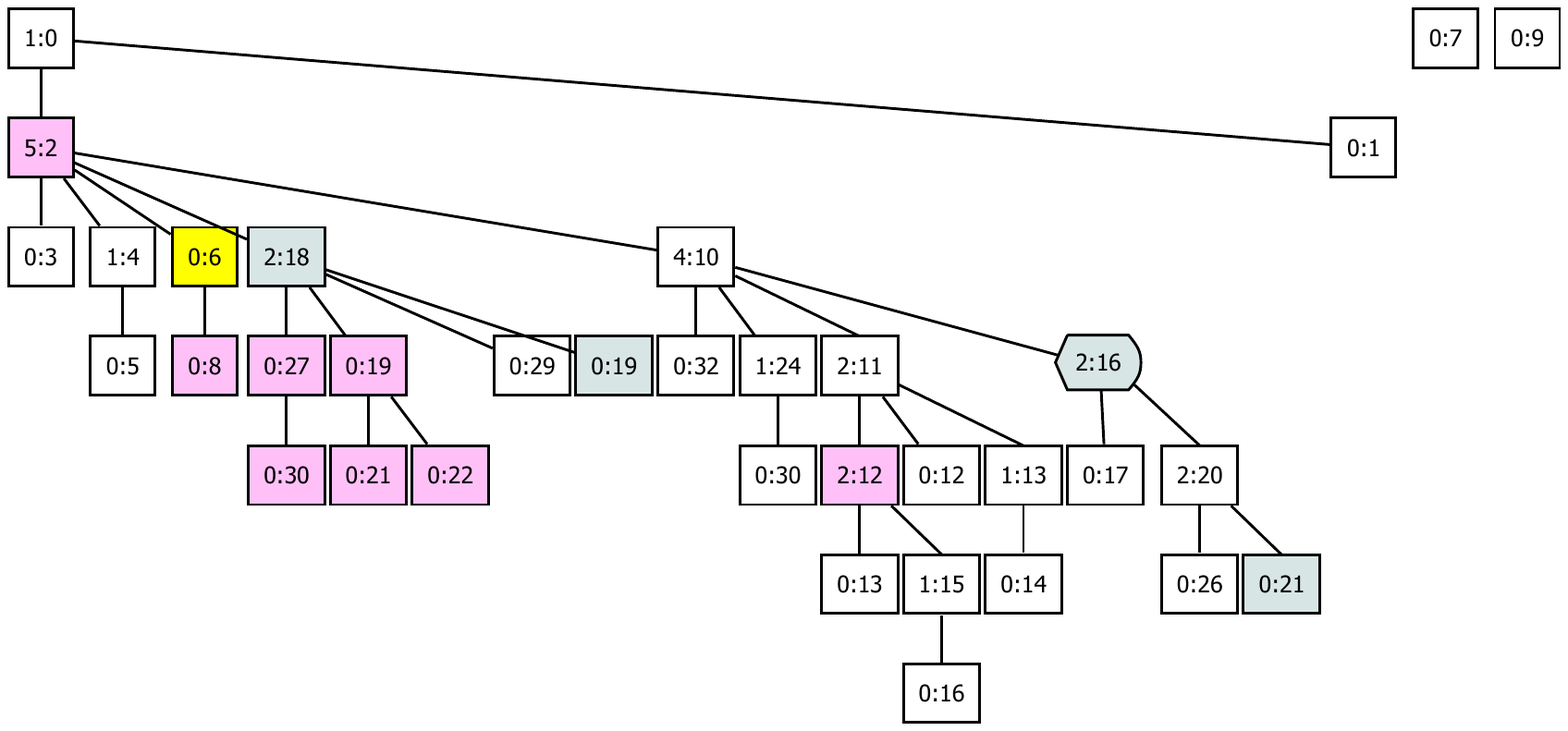}\rlap{\smash{\pdfsave\pdfrefximage\pdflastximage}}\pdfrestore
\kern 150mm\hss}
\kern-150mm
\hbox{fig 3: forest of heap ordered trees with ranks; yellow color denotes a misplaced outer child,}
\hbox{nonrectangular shape denotes a dangerous vertex}
\vss
}

Rank definition:\hfil\break
Let us define \nill\ pointers to be noncritical with rank -1. 
Let $w_0$ be last inner child of $v$ and $w_1$ second last, both could become \nill\ if such a child does not exist.\hfil\break
1. If $\rho_{w_0}>\rho_{w_1}+1$, and $w_0$ is noncritical then only $w_0$ is active and $r_v=\rho_{w_0}=r_{w_0}$.\hfil\break
2. If $\rho_{w_0}>\rho_{w_1}+1$ and $w_0$ is critical\numfootnote{We define $r_v=\rho_{w_0}=r_{w_0}+1$ temporarily for the analysis.}, we made $w_0$ outer and recompute rank again.\hfil\break 
3. Otherwise both $w_0$ and $w_1$ are active (or \nill) and $r_v=\rho_{w_0}+1$.

Rule 2 applies only temporarily, only the vertex whose rank is beeing recomputed could have this property.
Let us call vertices whose rank correspond to rule 1 dangerous while vertices whose rank correspond to rule 3 safe.

We will call application of either of rules 1., 2., 3. rank computation step. There can be several rank computation steps by rule 2 and at most one computation step by other rules. 

Vertices of rank 0 are safe.
\Insert\ creates safe vertex of rank 0 (with both $w_0$, $w_1$ \nill), rank updates by joins during first phase of \FindMin\ create safe roots as well. Ranks according to rule 1 could be achieved only by cuts and corresponding cascading rank recomputation process, rank of the vertex $v$ must have been at least $\rho_{w_0}+1$ when $v$ was safe. Therefore when vertex becomes dangerous it becomes simultaneously critical.
Later it could become outer (either during the same rank recomputation) or during another \Cut\ or \DeleteMin. 
Padovan heaps prevent creation of dangerous noncritical inner vertices, which would otherwise appear, by changing first phase of \FindMin, by making roots safe before we use their rank.
Details and cost of \MakeSafe, making one dangerous outer vertex safe, will be discussed later.
If we include safe test\numfootnote{$r_v>\rho_{w_0}$ test needs constant time, for a vertex $v$ of a positive rank, as $w_0$ is always the rightmost child},
and we call \MakeSafe\ if it fails, in the first phase of \FindMin, following invariants $(**)$, $(*{*}*)$, $(*{**}*)$ would hold:

$(**)$ noncritical inner vertices are dangerous only temporarily. 

Let inner children of $v$ from left to right in list are $i^v_0$, $i^v_1$, \dots, $i^v_{c_v+n_v-1}$. Then
$$(*{*}*)\kern 1cm\hbox{$\rho_{i^v_k}\ge k$.}$$
$$(*{**}*)\kern 1cm\hbox{$\rho_{i^v_{k+1}}\ge \rho_{i^v_{k}}+1$ for $k\ge0$.}$$

As we allow joins adding inner children only of safe (noncritical) vertices of the same rank, 
\FindMin\ maintains $(*{**}*)$. 
Nothing changes noncritical ranks, just vertices could be removed from the list of inner children, but it is compatible with the invariant. 
So the noncritical ranks of inner children are strongly increasing and $(*{*}*)$, $(*)$ become its trivial consequences 
(for $(*{*}*)\Rightarrow (*)$ consider $\rho_{w_1}=\rho_{i^v_{c_v+n_v-2}}\ge c_v+n_v-2$, both rules 1. and 3. lead to $r_v\ge c_v+n_v$).

There is one more detail we have not adressed yet.
When rank recomputation process decrements the rank of an innner vertex second time, we have to make the vertex outer.
But if it is not among the last two children, it has no quick access to the parent\numfootnote{we even cannot quickly detect if it has parent}.
This is why we divide outer vertices to placed and misplaced. Second phase of \FindMin\ creates placed outer children, while decrement of a critical child's rank makes the child outer misplaced.
We check whether the child is the rightmost or the secondrightmost.
If not, the rank recomputation ends, otherwise we have access to the parent and we recompute parent rank and place the child and possibly other misplaced outer children in the process. 
As the parent rank is not decremented when a child become outer misplaced, we neednot include misplaced vertices in $\Phi_1$. 
But we should prepay future placing of them so we have to count them in another potential $\Phi_5$. 
Let $m_v$ be the number of outer misplaced children of $v$ and let $p_v$ be the number of outer placed children of $v$. Than $\Phi_1=\sum p_v$, and $\Phi_5=\sum m_v$. Let us call placing step the operation placing a misplaced child during a rank recomputation.
Last potential we will need is $\Phi_6$, the number of dangerous vertices.

\vbox to 93mm{
\kern100mm
\hbox{\kern-2mm\pdfximage width 14cm {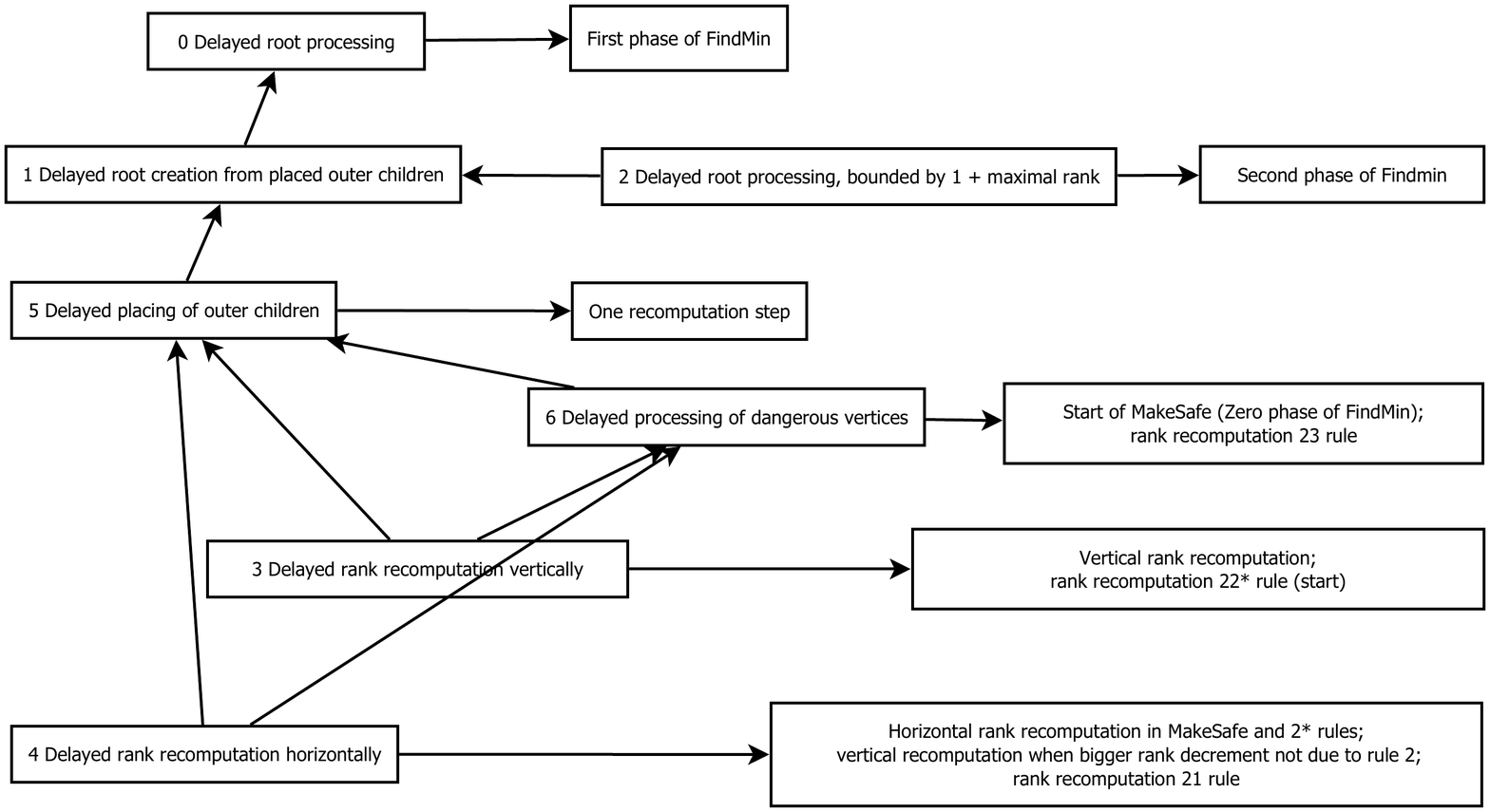}
\rlap{\smash{\pdfsave
\pdfrefximage\pdflastximage}}\pdfrestore
\hss}
\vss
\hbox{fig 4: payment schema of the amortized analysis}
\kern4mm
}
Now we have mentioned all the required potentials. 
The total potential according to which we do the amortised ananlysis is $t_0\Phi_0+t_1\Phi_1+t_2\Phi_2+t_3\Phi_3+t_4\Phi_4+t_5\Phi_5+t_6\Phi_6$ where $t_0$, $t_1$, $t_2$, $t_3$, $t_4$, $t_5$, $t_6$ are properly chosen constant times where $t_0\le t_1<t_2$, $t_1<t_5<t_6$, $t_5+t_6<t_3$, and $t_5+t_6<t_4$.
In the following we will discuss constants $t_i$ and we recapitulate the time analysis is correct with them. Analysis of rank restoration was not stated yet.

Remember that $\Phi_0=\tau$ be the number of trees in the heap, $\Phi_1=\sum p_v$ be the number of all outer placed children, $\Phi_2=\min\{\tau,1+\lfloor\log_\beta(n/c)\rfloor\}$, $\Phi_3=\sum c_v$ be the number of critical children, $\Phi_4=\sum_v r_v-(c_v+n_v)$ be the number of children cuts resp. making child outer not reflected in rank decrease yet, $\Phi_5=\sum m_v$ be the number of all outer misplaced children, and $\Phi_6$ be the number of all dangerous vertices.

Time $t_0$ suffices to work with a root during the first phase of \FindMin\ (one step in list traversal, testing the root is safe and callig \MakeSafe\ if not (\MakeSafe\ time is not paid from $t_0$), testing emptiness of the same root rank identifying place, putting a root to the same root rank identifying place, removing a root from the same root rank identifying place and joining two trees with the same root rank including removal of one of them from the list of roots). 

Time $t_1$ should be at least $t_0$ plus the time required to remove information about parent of outer child during \DeleteMin\ operation.
In fact, we maintain parent information only in the oldest outer child in case of root of rank 0, and the information is removed implicitly by joining the list of mimimum children with the list of tree roots, so no additional time is needed and therefore $t_1=t_0$. 

Time $t_2$ should be at least $t_1$ plus the time required in the second phase of \FindMin\ operation per remaining tree. 
We consider emptying the same root rank identifying places to be part of the second phase in this analysis so time for traversing one root and emptying coresponding place is incorporated in $t_2-t_1$ as well as time to join two roots including removal of one of them from the list and 2 steps to traverse a root during linking traversals of the list of trees. 

Time $t_5$ should be at least $t_1$ plus the time to cut an outer misplaced vertex from the children list, and time to place it as outer placed vertex to the list start, the parent is known when we need to access the list start. It includes the time to test the vertex is outer misplaced.

Time $t_6$ should be at least $t_5$ plus the time to change the status of rightmost child from inner (critical or noncritical) to outer misplaced, 
and time for one rank recomputation step by rule 2 and one by other rule and the time to stop the while loop for \MakeSafe\ if the parent becomes safe.

Time $t_3$ should be at least $t_5+t_6$ plus the time to change the status from critical to outer misplaced, and time to test that the vertex is critical and among the two rightmost children and step to the vertex parent (in that case). It also covers one rank recomputation step by rule 2 and one by other rule. 

Time $t_4$ should be at least $t_5+t_6$ plus the time to change the status from noncritical to outer misplaced, and time to test that the vertex is noncritical and among the two rightmost children and test the two rank differences exceeds 1 and step to the vertex parent in that case. It also covers one rank recomputation step by rule 2 and one by other rule and time for one step in while loop for \MakeSafe\ if the parent remains dangerous.

Each placing step is paid by $t_5-t_1$, therefore cost of a rank computation step is constant. 
Rank recomputation of a vertex $v$ includes at most one step according to rules 1 or 3. 
$\Phi_3$ could be increased once if the affected vertex changes from noncritical to critical child during the recomputation.
It could include several steps according to rule 2.
Each recomputation step according to rule 2 decreases $\rho_{w_0}$ by at least 2 and $c_v$ by 1 and preserves $n_v$. 
It decreases $\Phi_3$ (not counting criticaity of recomputed vertex) by 1 and except the last $\rho_v$ is decreased to next $\rho_{w_0}$ so $\Phi_4$ decreases by at least 1. The last step ends with $\rho_v=\rho_{w_0}+1$ so $\Phi_4$ does not increase.  

Now we can return to \MakeSafe. What is its cost? And how it could be coded?
While $v$ is dangerous, we could make $w_0$ outer misplaced and recompute the $v$'s rank\numfootnote{We could localise all vertices which should become outer this way by searching $v$'s children leftward, till we found two inner children whose noncritical rank differ by at most 1 or inner child of rank 0 is reached or we reach outer placed child. Traversed inner children and missplaced children are put to the left end of the children list as placed children.}.
Time to making the vertex safe would be proportional to the number of vertices traversed.
Outer misplaced vertices already have prepaid $t_5-t_1$ for its change to outer placed and move to the start of the list in $\Phi_5$.
Let $i^v_{c_v+n_v-k}$, \dots, $i^v_{c_v+n_v-1}$ be inner children among traversed vertices. 
Except one step all recurent steps are paid by $\Phi_4$.  
The step unpaid this way does not increase $\Phi_4$. As we call \MakeSafe\ only at roots, $\Phi_3$ is not increased by it. And no new critical vertex is introduced. $\Phi_6$ pays for constant time and first increase of $\Phi_5$ so full \MakeSafe\ is prepaid.

\Meld\ operation works in scenario when we use several heaps at the same time.
Meld unions sets of two of them.
The operation just connects lists of tree roots (and removes one dummy head).
In analysis we should sum correspondning potentials.
The final sum could only decrease in case $\Phi_2$ exceeded it's upper bound. 
So the cost for the \Meld\ operation with respect to the unified potential is constant.

\Insert\ could be implemented as creation of the heap with only one tree with only one vertex followed by \Meld\ with the original heap\numfootnote{Better is to omit temporary creation of a dummy head.}. As the potential of the created heap is constant, the cost of \Insert\ operation is therefore constant.

Let us bound \FindMin\ cost except the cost of \MakeSafe{}s.
\FindMin\ ends with $\Phi_0=\Phi_2=1$, $\Phi_1$ is usually increased by \FindMin, $\Phi_3$ is not changed. $\Phi_4$, $\Phi_5$, and $\Phi_6$ are not changed except by \MakeSafe{}s. Constants $t_0$, $t_1$ and $t_2$ were chosen such that it's time and increase of $\Phi_1$ is fully paid by decrease of $\Phi_0$ and $\Phi_2$, so the cost of \FindMin\ is constant.

\DeleteMin\ on minumum $v$ removes in constant time the tree with $v$ from the original heap (it was the only tree of the heap). 
It creates in constant time heap from $v$'s children and melds the two heaps\numfootnote{Again omitting temporary creation of a dummy head}. 
It increases $\Phi_0$ by $\iota_v-1=p_v+m_v+c_v+n_v-1\le p_v+m_v+r_v-1\le p_v+m_v+\lfloor\log_\beta(n/c)\rfloor$ and as $\Phi_1$ is decreased by $p_v$ and $\Phi_5$ by $m_v$ the cost is bounded by $O(1)+(\lfloor\log_\beta(n/c)\rfloor)(t_0+t_2)-p_v(t_1-t_0)-m_v(t_5-t_0)\in O(\log n)$ (we count with maximal possible increase of $\Phi_2$). Possible decrements of $\Phi_3$, $\Phi_4$, and $\Phi_6$ could only reduce the cost.

\Decrement\ could be implemented by \Cut\ followed by cascading rank update followed by decrement of the key in the newly forced root. The decrement of the key in the root requires constant time and does not change potential, so we have to prove constant cost of \Cut\ followed by cascading rank updates, to finish its analysis.

\Delete\ could be implemented by \Cut\ followed by cascading rank update followed by delete of the newly forced root. The delete of the newly forced root analysis would be same as analysis of \DeleteMin\ (except now there could be more trees in the heap), so constant cost of \Cut\ followed by cascading rank updates would be sufficient to prove the cost is bounded by $O(1)+(\lfloor\log_\beta(n/c)\rfloor)(t_0+t_2)\in O(\log n)$.

So, finally, what happens during \Cut($v$) and following cascading rank updates? 
We detect by $v\to\Right\to\Right\to\Left\to\Left\not=v$ the vertex is among the last two vertices on its list. If not, the update ends by cutting $v$.
If it is, we check by $v\to\Right\to\Left\not=v$ the vertex is last. In both cases we know the vertex parent $p$ and we could test if it is not dummy head by 
$p\to\Right\not=p$. Otherwise we end as $v$ was root. In the last case the rank recomputation starts at the parent $p$ after $v$ is cut.
Cut of nonroot increases $\Phi_0$ by 1 and could increase $\Phi_2$ by 1 as well. Depending on type of cut vertex it could decrement either $\Phi_1$, 
$\Phi_3$, $\Phi_5$ or none of them. It could increase $\Phi_4$ by 1. If $p$ becomes dangerous, $\Phi_6$ is increased by 1 as well.
It takes constant time and makes at most constant change to the potential so it's cost is constant. Cut of root acts similarly as we usually do not detect the vertex is root, but there is no change in the potential in that case.
What we should analyse is the following rank recomputation process.

We start by seeking the active inner children. While the last child is outer misplaced, we cut it and reinsert it as outer placed on the start of children list\numfootnote{We could change them to placed and move them all at once by at most 4 pointer changes when inner child is found.}.
$\Phi_5$ is decremented by 1, $\Phi_1$ incremented by 1 and $t_5-t_1$ pays for the required time.
When the last child is inner, we seek for the second last the same way\numfootnote{Thanks to $(*{*}*)$ we could stop seeking for second inner child when the second child is outer misplaced. We will not found $w_1$, but we would know rule 3 would not apply.}.
Whenever the considered child is outer placed, we could stop the seeking. 
As result of the process we got those of $w_0$ and $w_1$ whose are important in constant cost, and we could detect which rule of the rank definition would apply. 
If it is rule 2, we have to make the critical vertex outer misplaced and continue the seeking. 
Cost of the rank recomputation of one vertex is time for one step by rule 2 and one step by another rule.
When the rule 2 applies $k+1$ times, the rank droped by at least $2k+1$ so $\Phi_4$ was decremented by at least $k$ and
$k$ uses of rule 2 are paid by\numfootnote{Decrements of $\Phi_3$ remain in reserve for other use} $k(t_4-t_5)$.

To simplify the codding we could stop cascading rank recomputation when $p$ is dummy head or rank of $p$ does not change or when $p$ is not among the last two children. Calling the rank recomputation in the case rank cannot change would cost us at most constant as after recomputation of the rank the process terminates and it's worth the simplification. 

The cost of the last rank recomputation is constant as well as the second last.
What should be analysed carefully are cases when the rank recomputation does not stop on the parent $g$ of $p$.
This means $p$ must have been active inner child of $g$.

If $p$ was a critical active child and its rank decrements by a positive amount, it becomes outer misplaced and $\Phi_3$ derements by 1, $\Phi_5$ increments by 1, and $\Phi_4$ cannnot increase. If $g$ becomes dangerous, $\Phi_6$ is increment by 1. $\Phi_3$ pays $(t_3-t_5-t_6)$ for the continuation.

If $p$ was a noncritical active child and becomes outer misplaced, it's rank was decremented by at least 2.
If rule 2 was applied on $p$, we can pay $(t_3-t_5-t_6)$ for the continuation by a decrement of $\Phi_3$.
Otherwise $p$ lost exactly one inner child (could become outer) and $r_p-c_p-n_p$ (and therefore $\Phi_4$) was decremented by at least 1.
$\Phi_5$ is incremented by 1 and $\Phi_6$ could increment by 1. $\Phi_4$ pays $t_4-t_5-t_6$ for the continuation.

The last case is when the active children of $g$ remain, but the rule 2 applies to $g$.
This could happen when $p$ was noncritical and the only active child of a dangerous $g$, and $p$ become critical. 
Thanks to $(**)$ we know $g$ cannot be noncritical inner vertex at the start of the \Cut.
If $g$ was outer, rank recomputation would stop and the final cost is not important constant.
If $g$ was critical, we would use $\Phi_3$ to pay continuation in it's predecessors.
We just should discuss resources to pay for rank recomputation of $g$. 
It depends on the second applied rule.
If rule 2 was applied at least twice, we have $\Phi_3$ decrement to pay from.
If rule 3 applies, $g$ changed from danger to safe and we have $\Phi_6$ decrement to pay from.
Finally if rule 1 applies, $r_g$ was decremented by at least 2 and just one inner child become outer, therefore we have $\Phi_4$ decrement to pay from.

As the last two steps of cascading rank consolidation have constant cost and the other steps are fully paid by the potential decrease, the total rank recomputation cost is constant.

\section{Concluding Remarks}
The Fibonacci heaps don't require define outer children, but if we sacrifice a bit to allow them, we can use outer children even in Fibonacci heaps.
The $\Phi_1$ would coincide with the original definition and there is no problem in stopping cascading rank consolidation in outer children (the rank invariant would hold in both cases with the same $\beta$). It will better correspond to standard Fibonacci heaps with the only difference some roots of standard Fibonacci heaps would be hidden as outer children of other vertices.

All mentioned variants of Fibonacci heaps and Padovan heaps respecting the superexpensive comparison principle could be compared to standard Fibonacciho heaps by a following competition: Supporter of one structure defines sequence of calls to the methods on empty initial heap. The sequence is run on both implementations and the total times of the implementations are divided, and the result defines the gain.

I claim that in such a game our variants would lose by at most constant when standard implementation supporter choses the sequence and
on the contrary there exists a sequence where standard implementation loses by $\Omega(\log n)$ where $n$ is length of the sequence
(prefix of sequence generated for increasing i by adding:$\{x_i=\Insert(-i)$, $x_{i+1}=Insert(-(i+1))$, $m=FindMin()$, $\DeleteMin()\}$).
The complicated second phase presented in \FindMin\ implementation details guarantees there would be at most two outer children of the minimum after each \FindMin\ and others would hide forever on deeper levels of the tree. 
This prevents creation of rank 4 so the average time remains constant.
Actually you can see there is no \Decrement\ in the used sequence, so the shown inefficiency doesn't need it.

There is counterargument against this competition. As two standard Fibonacci heap implementations that differ only in small datail defining order of pairing the roots of the same rank could create sequences whose are good against the other. When creating heap of size $2^k+1$ by inserts in propper order they could force their implementation to create a heap where minimum is lone in its tree while the other vertices are in one huge tree. On the opoosite it's highly probable the minimum is in the huge tree in the other implementation. Following \DeleteMin\ and reinsert the deleted key could repeat the situation with costs of 1 and $k$. (To prevent the repetition the heap should link the last inserted vertex last, what actually our implementation does as we insert them to the end of the list.)

The claim is not proven and we could probably improve the argument by defining fair competitions, but this is definitely out of scope of this paper.

\vbox to 80mm{
\hbox{\hss\kern 2cm\pdfximage width 9cm {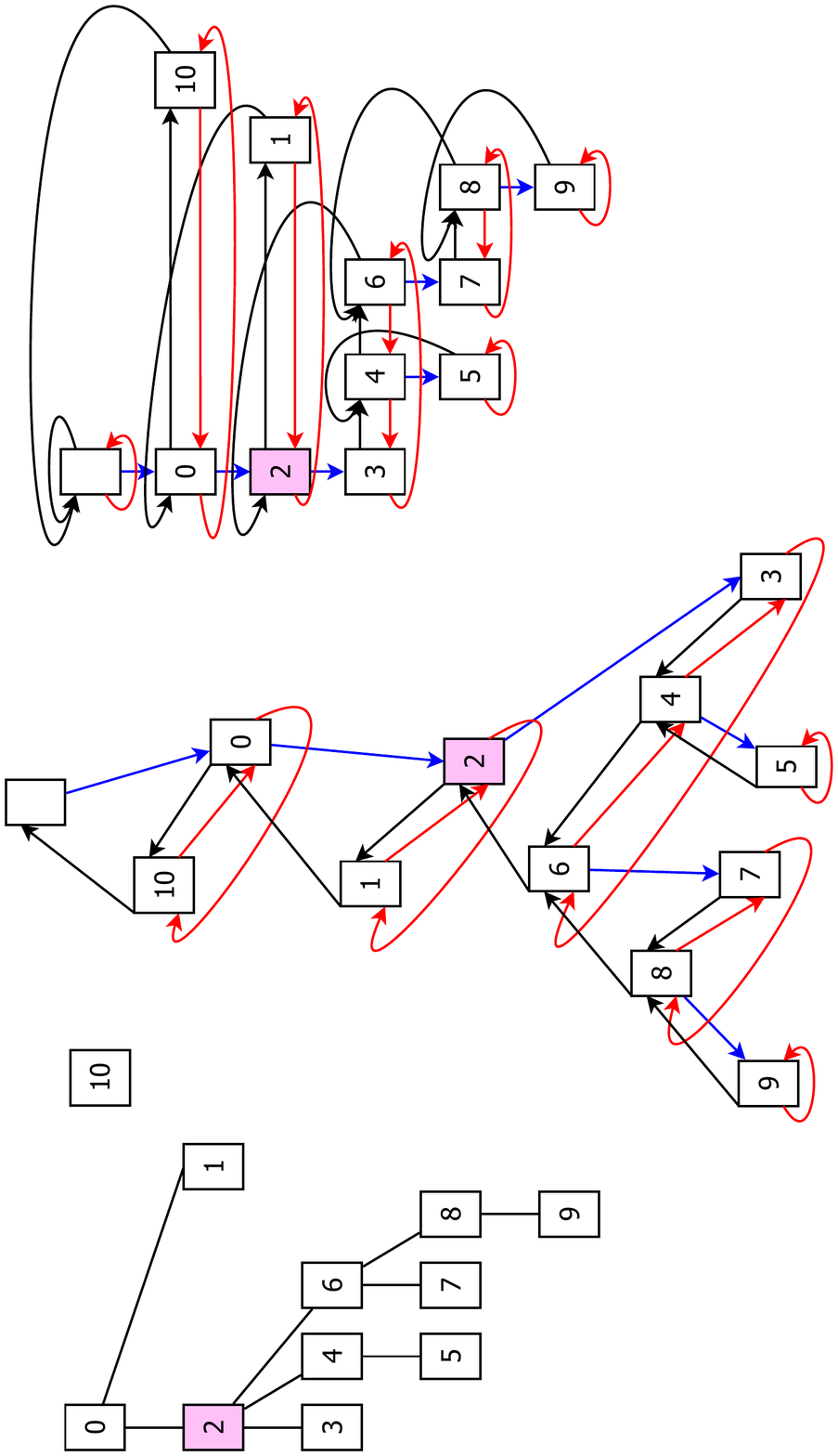}\rlap{\smash{\pdfsave\pdfsetmatrix{0 -1 1 0}\pdfrefximage\pdflastximage}}\pdfrestore
\kern5cm}
\vfil
\hbox{fig 5: forest of heap ordered trees, it's representation in rank pairing heaps (one tree}
\hbox{variant), and in Padovan heaps (isomorphism except the dummy root)}
\kern5mm
}

Rank pairing heaps \cite{RankPairingHeaps} of B. Haeupler, S. Sen, R. E. Tarjan use 3 pointers per node as well as Padovan heaps, so they are alternative solution to the same problem. They are another implementation of Fibonacci heaps. 
Actually at chapter 6 they mention unfair links, so they follow the superexpensive principle there.
The trees are isomorphic, but the balancig differs. Our argumentation is strictly based on the principle, while their just confirms the principle is worthwhile.

\section{Summary}
We have shown one pointer could be saved in Fibonacci heaps in rather conservative extension of their variant.
Actually we gain assymptotically same amortized time bounds for operations, but the rank invariant leads to renaming the heaps to Padovan as the base of logarithm for \DeleteMin\ bound has changed. 
We have introduced the superexpensive comparison principle and showed the Fibonacci heaps could be implemented according to it, and Padovan heaps extension follows the principle as well.

Real computers do not correspond to pointer machine model presented here. 
When we consider cache hierarchies, we will find these structures not well optimalised to block reads (cache misses). 
But this was not addressed by the article.


\bibliography{padovan}
\end{document}